\documentclass[12pt,preprint]{aastex}
\usepackage{amsmath, amsthm, amssymb}

\slugcomment{To appear in ApJ}

\shorttitle{Planarity of Globular Clusters}
\shortauthors{Keller, Mackey \& Da Costa}

\begin{document}


\title{The globular cluster system of the Milky Way:\\accretion in a cosmological context}

\author{Stefan C.\ Keller}

\author{Dougal Mackey}

\and

\author{Gary S.\ Da Costa}
\affil{RSAA, Australian National University, Weston Creek, ACT 2611, Australia}

\begin{abstract}
We examine the significance of a planar arrangement in the spatial distribution of the Milky Way's globular clusters (GCs). We find that, when separated on the basis of horizontal branch morphology and metallicity, the outer-most canonical young halo GC sample (at galactocentric radii in excess of 10 kpc) exhibit an anisotropic distribution that may be equated to a plane ($24 \pm 4$) kpc thick (rms) and inclined at $8^{\circ} \pm 5^{\circ}$ to the polar axis of the Milky Way disk. To quantify the significance of this plane we determine the fraction of times that an isotropic distribution replicates the observed distribution in Monte-Carlo trials. The plane is found to remain significant at the $>95$\% level outside a galactocentric radius of 10 kpc, inside this radius the spatial distribution is apparently isotropic. In contrast, the spatial distribution of the old halo sample outside 10 kpc is well matched by an isotropic distribution. The plane described by the outer young halo globular clusters is indistinguishable in orientation from that presented by the satellite galaxies of the Milky Way. Simulations have shown that the planar arrangement of satellites can arise as filaments of the surrounding large scale structure feed into the Milky Way's potential. We therefore propose that our results are direct observational evidence for the accreted origin of the outer young halo globular cluster population. This conclusion confirms numerous lines of evidence that have similarly indicated an accreted origin for this set of clusters from the inferred cluster properties.
\end{abstract}





\section{Introduction}

It has long been noted that the most luminous satellite galaxies of the Milky Way (MW) exhibit an asymmetric spatial distribution as seen on the sky \citep{Lynden-Bell76, Kunkel76, Lynden-Bell82}. \citet{Kroupa05} demonstrated that the spatial distribution of the 11 most luminous MW satellites is inconsistent with an isotropic distribution in 99.5\% of random realisations. Rather, these satellite galaxies describe a planar structure that extends 254 kpc from the MW and possesses a scale height of $\sim20$ kpc; the MW plane of satellites (PoS) \citep{Metz07, Metz08}\footnote{We do not use the term `disk of satellites' of these authors as the term `disk' implies a circular or rotationally-supported configuration in three dimensions that is not constrained in our discussion to follow.}. \citet{Metz09a} extends this work to include the newly discovered ultra-faint dwarf galaxies with minimal change to the determined spatial parameters of this plane.

The PoS so described is seen to be within $\sim 8^{\circ}$ of polar to the rotation axis of the MW. The Andromeda galaxy (M31) is the only other extensive satellite system for which we can describe the 3-d structure with meaningful accuracy. A similar degree of planarity is evident around M31 \citep{Koch06}. \citet{Metz09a} updated this analysis to include increased numbers of M31 satellites and find that the plane of satellites around M31 is inclined at 59$^{\circ}$ with respect to the disk (it should be noted that the number of M31 satellites has increased by $\sim50$\% since the work of \citet{Metz09} however; see \citet{Martin09}, \citet{Richardson11}). 

The configuration of satellites about their host galaxy rose to prominence with the study of \citet{Holmberg69} who proposed that satellites are preferentially found near the poles of the host galaxy disk (the `Holmberg effect'). Subsequent studies based on extensive redshift surveys have not confirmed this effect \citep{Sales09, Bailin08}. When averaged over all host-satellite pairs, the satellites of relatively isolated galaxies (akin to the MW and M31) have a tendency to be found near the major axis of their host \citep{Brainerd05,Agustsson10}. The  findings that the satellite planes of the MW and M31 are highly inclined are therefore unexpected.

Recently, \citet{Kroupa10} have argued that the existence of the MW PoS is one of a number of critical challenges for Lambda CDM ($\Lambda$CDM) cosmology. \citeauthor{Kroupa10} propose the MW satellite system is a system of tidal dwarf galaxies (TDGs) that resulted from a major `wet' merger that gave rise to the galactic Bulge. However, \citet{Libeskind10} and \citet{Lovell10} contend that the observed spatial distribution and possible rotating configuration of the MW PoS is not an insurmountable problem for $\Lambda$CDM and can arise naturally from hierarchical structure growth. In their simulations the MW satellite system properties can be frequently matched when accretion occurs preferentially along filaments of cosmic dark matter (DM) structure. Issues concerning cosmogony aside, the MW PoS is accepted to comprise a system that has resulted from accretion to the MW halo.

It was the seminal work of \citet{SZ} that utilized the horizontal branch (HB) morphology of the outer halo globular clusters (GCs), together with the absence of an abundance gradient within the outer halo GCs, to propose that accretion onto the early MW was substantial, and in the case of the outer halo, the dominant formation mechanism. Subsequently, \citet{Zinn85, Zinn93} and more recently, \citet{Mackey04} present a reanalysis of the Searle \& Zinn result. In these previous studies the MW GC population is split on the basis of HB morphology and metallicity into three sub-groups: an old halo GC population which have blue HB morphology at given [Fe/H] and ages $\sim$ 13 Gyr; disk/bulge GCs of relatively high metallicity and exclusively red HB morphology; and a 'young halo' (YH) population of GCs with comparatively red HB morphology at given [Fe/H]. 

The YH GCs, as defined by \citeauthor{Mackey04}, span a distance range far in excess of the other populations: from 2 -- 120 kpc, as distinct from, for instance, the old halo population where the majority reside within 30 kpc. In terms of kinematics, the YH GC system shows no evidence for net rotation, while the old GC system shows prograde rotation. \citeauthor{Mackey04} demonstrate the young system resembles the GC systems of the Large and Small Magellanic Clouds (LMC and SMC), Fornax and Sagittarius dwarf galaxies in terms of metallicity range, age and HB morphology. Furthermore, the YH GC population possess generally more extended structures than their old halo counterparts. The distribution of core radii for the YH system is seen to match that of the above mentioned four external dwarf galaxies, and distinct from the core and half-light radius distribution of the old halo GC subsystem to high significance \citep{Mackey04}. Based on these above similarities between the YH GC subsystem and the external GC systems, \citet{Mackey04} propose that the YH GCs are of extra-galactic origin (in accordance with the result of \citeauthor{SZ} \citeyear{SZ}), whereas the majority (but not all) of the old halo and bulge/disk GCs are proposed to have formed in-situ under some form of dissipative collapse as originally envisaged by \citet{ELS}. 

The study of \citet{Yoon02} presented evidence for a planar arrangement of a set of seven Galactic GCs (GGCs). These seven clusters were selected based on the RR Lyrae Oosterhoff dichotomy (as Oosterhoff Type-2 clusters) and metallicity (the lowest metallicity clusters in the Type-2 group). \citet{Yoon02} propose these clusters have been accreted from an external galaxy. However as pointed out by \citet{Catelan09}, the present-day Milky Way satellites possess GC systems that do not match the Oosterhoff dichotomy and the distribution of GGCs in the HB morphology-metallicity plane. This discrepancy with present-day satellite galaxies aside, the studies of \citet{Marin-Franch09}, \citet{Dotter10}, and \citet{Dotter11} show that the age-metallicity relation of the GGCs possesses two branches -- one with near constant old age of $\sim 13$ Gyrs and another branch to significantly younger ages with increased [Fe/H]. The younger branch of GGCs is found to be dominated by GCs associated with the Sagittarius and Canis Major dwarf galaxies. This leads \citet{Forbes10} to conclude that approximately one quarter of the GGC system has resulted from accretion.

In this study, we show that the YH GCs of the MW (as defined by a combination of age estimators) are confined to a plane that is indistinguishable from the MW PoS. The YH GCs act as tracers of the dwarf galaxies that have been disrupted in the formation of the outer halo. The consistency in spatial alignment of early accretion events to the alignment of current satellites suggests a preferred orientation for accretion from the local large scale structure that has remained consistent for a large fraction of a Hubble time.

\section{Partition of the globular cluster population by relative cluster age}
\label{Section:partitionGCs}

When excellent self-consistent photometry exists, such as in the study of \citet{Marin-Franch09}, it is possible to derive accurate relative ages even amongst clusters of advanced age. These authors compare the luminosity of the main-sequence turn-off to that of the lower main-sequence and hence express the relative age for a sample of 64 nearby GCs. \citeauthor{Marin-Franch09} finds evidence for two groups of GCs: an old GC population with an age dispersion of $\sim5$\% and no age-metallicity relation, and a young GC population with an age-metallicity relation similar to that defined by the GCs associated with the Sagittarius dwarf galaxy. 

\begin{figure*}
\begin{center}
\includegraphics[scale=1, angle=0]{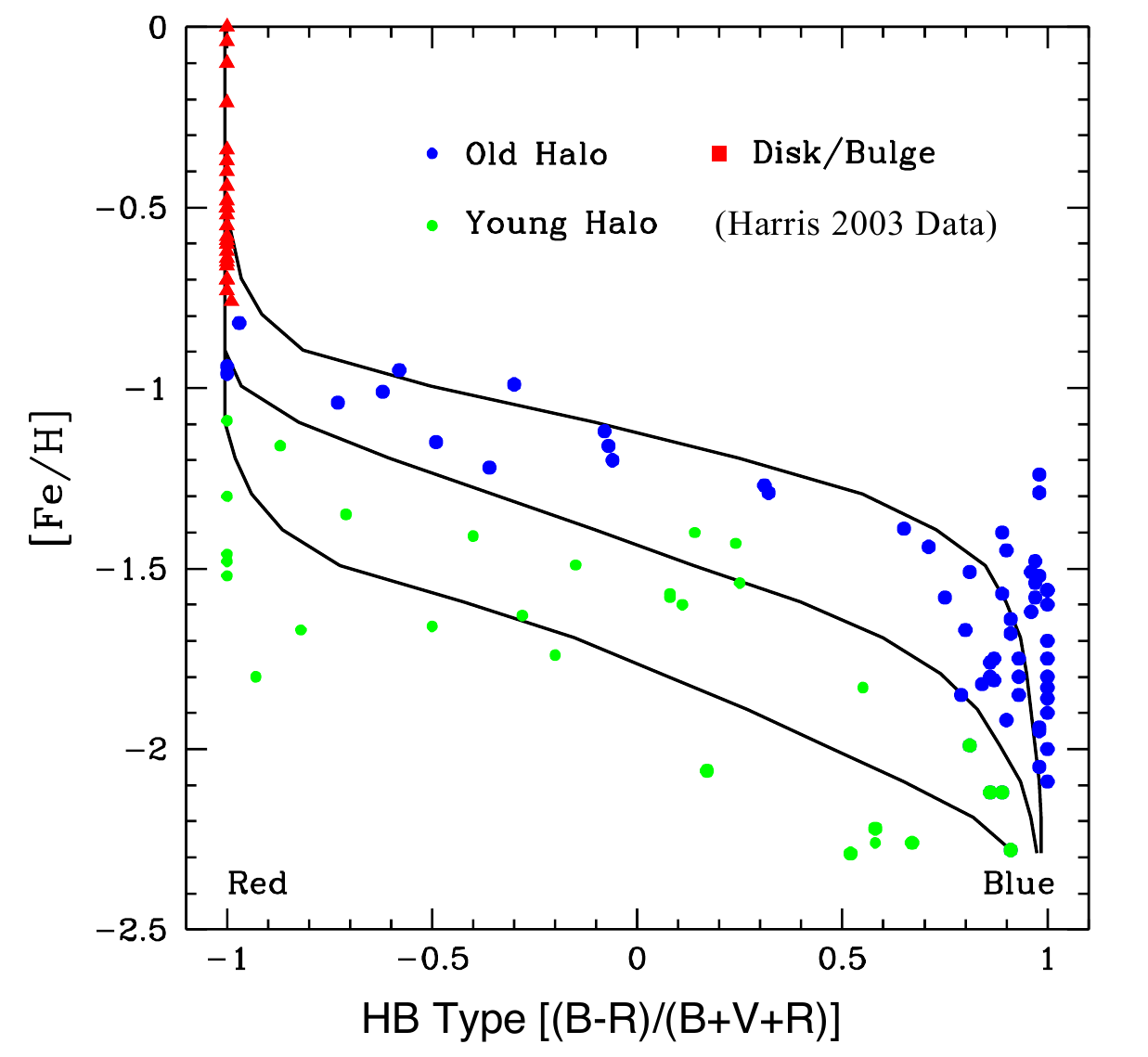}
\caption{HB-type versus metallicity diagram for the 114 Galactic globular clusters with suitable measurements in the 2003 \& 2010 versions of the Harris (1996) catalogue. The clusters have been divided into three subsystems, as labelled, according to the criteria set out in the text. The overplotted isochrones are from \citet{Rey01}. The two lower isochrones are, respectively, 1.1 Gyr and 2.2 Gyr younger than the top isochrone.}\label{figure:GC_Groups}
\end{center}
\end{figure*}

The age determination method of \citet{Zinn93} relies on a relation between age, metallicity, and HB morphology to resolve relative cluster age differences. \citeauthor{Zinn93}  defined a fiducial sequence for inner halo clusters ($R_{gc} < 6$ kpc) in a HB-type\footnote{HBR = $ \left[\frac{(B-R)}{(B+V+R)} \right] $ where $B$ is the number of stars bluer than the instability strip, $V$ is the number of stars within the instability strip, and $R$ is the number of stars to the red of the instability strip.} against [Fe/H] diagram and then measured the difference between the HB-type of a given cluster and that of the fiducial sequence. Those with a difference in HB-type $>$ $-0.4$ were classified by \citet{Zinn93} as young clusters, the remainder old clusters. \citet{Mackey04} has revisited the issue of relative age partition. Rather than using spatial information in the derivation of their fiducial these authors utilise the model evolutionary isochrones of \citet{Rey01}. The isochrone corresponding to the canonical ancient GC population traces the those clusters with the bluest HB-type at a given [Fe/H] (see Figure \ref{figure:GC_Groups}). In the study of \citet{Mackey04} the young halo GC population is defined as those clusters with a difference in HB-type in excess of $-0.3$ from that of the fiducial at a corresponding [Fe/H]. This corresponds to a minimum age difference of around -0.6 Gyr at HB-type = 0. It should be noted that there is an important limitation the application of this technique for estimating the relative ages between clusters. At the extremes of HB-type the isochrones become degenerate within observational uncertainties. We note that parameters other than age can also be responsible for the location of a cluster in these groupings. For example, in the case of NGC 2808 the extreme blue horizontal branch morphology is likely driven by He abundance variations. 

In the study of \citet{Yoon02} relative age differences of around 1 Gyr are inferred from an examination of the mean period of type-ab RR Lyraes ($\langle P_{ab} \rangle$) with cluster metallicity. Oosterhoff Group I clusters possess $\langle P_{ab} \rangle \sim 0.55$ days and are more metal-rich than those of Group II with $\langle P_{ab} \rangle \sim 0.65$ days. \citet{Yoon02} demonstrate that Group II clusters may be further split into `old' and `young' clusters (Groups II-a and II-b, respectively). The benefit of this method is that we can continue our classification to the blue extreme of HB-type. A limitation on the method is that the $\langle P_{ab} \rangle$ of many clusters is either limited by intrinsically low RR-ab numbers or the absence of studies in the literature \citep[see][]{Catelan09}.

Our partitioning of the GGC population will rely on a combination of the latter two methods. In the regime where both methods are viable (that is HB-type $< 0.8$, and $\langle P_{ab} \rangle > 0.60$ d) there are six objects in common: both methods classify five as `young', the exception is the GC Rup 106 for which the ($\langle P_{ab} \rangle$, [Fe/H]) method would suggest `old' and the (HB-type, [Fe/H]) method would suggest `young'. Examination of the MSTO in the CMD demonstrates that it is indeed younger than most GCs \citep{Marin-Franch09}. For the analysis that follows we will use the following criteria:
\begin{center}
\begin{tabbing}
{\texttt{if}} \= ((HBR - HBR$_{fiducial} < -0.3$) {\texttt{and}} HBR$ < 0.8$) \\
\>{\texttt{or}} ([Fe/H]$< -1.9$ {\texttt{and}} $\langle P_{ab} \rangle > 0.6$ days) \\
\>{\texttt{then}} classification = young halo cluster\\
{\texttt{else if }} ([Fe/H] $>-0.8$)\\
\>{\texttt{then}} classification = disk/bulge cluster\\
{\texttt{else}} classification = old halo cluster
\end{tabbing}
\end{center}

Graphically, our partition of the GC population can be seen in Figure \ref{figure:GC_Groups}. We take the [Fe/H], and in the analysis to follow, the distances and positions of GCs from the compilation of \citet{Harris96} (2010 edition). The HBR is taken from the 2003 edition of the \citeauthor{Harris96} catalog. Koposov 1 \& 2 are not included here as they lack sufficient dat to constrain their HB-type. The pulsation properties of the clusters are taken from the compilation of \citet{Catelan09}. The three GC subgroups young halo (YH), old halo (OH), and disk/bulge (DB), as defined above contain 30, 56, and 28 clusters respectively. Compared to the work of MG04, our selection criteria sees the clusters NGC 5024, 6341, and 7099 included in our YH grouping on the basis of the pulsation properties of their RR Lyrae populations, whereas these clusters were previously assigned to the OH grouping. 

\section{The spatial distribution of globular clusters}
\label{section:spatial_dist}

We now investigate how isotropic the spatial distributions of the three GC populations are. We utilise the method described in detail in \citet{Metz07}. In brief, we utilise a least squares technique to define the plane that best matches the spatial distribution and determine the normal vector to this plane. Since the least squares method is unweighted we account for distance uncertainties by repeat realisations of the system with objects randomly shifted along their lines of sight according to a normal distribution function in which the variance matches distance uncertainty to each object. This provides an estimation of the robustness of a disk-like population. If the GC sample is not distributed in a well-defined plane then repeat realisations of the sample to account for distance uncertainties will show a large scatter in the orientation of the normal vector. A large number of realisations ($10^4$) of the system are made. 

This results in a set of $m$ unit normal vectors ${\bf{\hat{n}}}$ for which $(\hat{x}_i, \hat{y}_i, \hat{z}_i)$ are the cartesian components. We then define the matrix {\bf{M}} as

\[ {\bf{M}} = \left( \begin{array}{ccc}
\hat{x}_1 & \hat{y}_1 & \hat{z}_1 \\
\hat{x}_2 & \hat{y}_2 & \hat{z}_2 \\
\vdots & \vdots & \vdots \\
\hat{x}_m & \hat{y}_m &  \hat{z}_m\end{array} \right)\] 

and perform an eigenvalue analysis of the matrix {\bf{T}}={\bf{M}}$^{\rm{T}}${\bf{M}}. The eigenvector that corresponds to the largest eigenvalue of {\bf{T}} is the principal axis of the set of input vectors. This principal axis defines the orientation of spatial anisotropy in the input GC sample. There are three eigenvalues of {\bf{T}}, $\tau_1 < \tau_2 < \tau_3$, where $\tau_3$ is the eigenvalue of the principal axis. In the case of a uniform spherical distribution $\tau_1 = \tau_2 = \tau_3 = 1/3$. Following the formalism of \citet[][their section 3.4]{Fisher87}, the following parameters derived from the eigenvalues provide us with a measure of how isotropic the distribution is. We define a shape parameter $\gamma$:

\[  
\gamma = \frac{\ln(\tau_3/\tau_2)}{\ln(\tau_2/\tau_1)}
\]

and a strength parameter $\zeta$:

\[  
\zeta = {\ln(\tau_3/\tau_1)}
\]

The parameters $\gamma$ and $\zeta$ enable the classification of the spherical distribution of normal vectors. For example, a high value of $\gamma$ (say 3 or more) indicates a clustered distribution of normals on the sky. When combined with a high value for $\zeta$, this indicates a strongly condensed distribution. A low value for $\gamma$ indicates the presence of a girdled distribution, i.e.\ a distribution that forms a band around the sky. $\zeta$ then describes how strongly confined the distribution is to this girdle.


We now consider the distributions of our three GC groups. We compare the spherical distribution of the normal vectors to the best-fitting plane with those arising from simulations of an isotropically distributed model GC population. The isotropic GC model is constructed from a density power law of index $-3.5$, which is appropriate for the overall MW GC population \citep{Parmentier05}. In galactic coordinates this model population is uniformly distributed. The number of simulated GCs is matched to the observed number of GCs in each age group. 

\begin{figure*}
\begin{center}
\includegraphics[scale=0.8, angle=0]{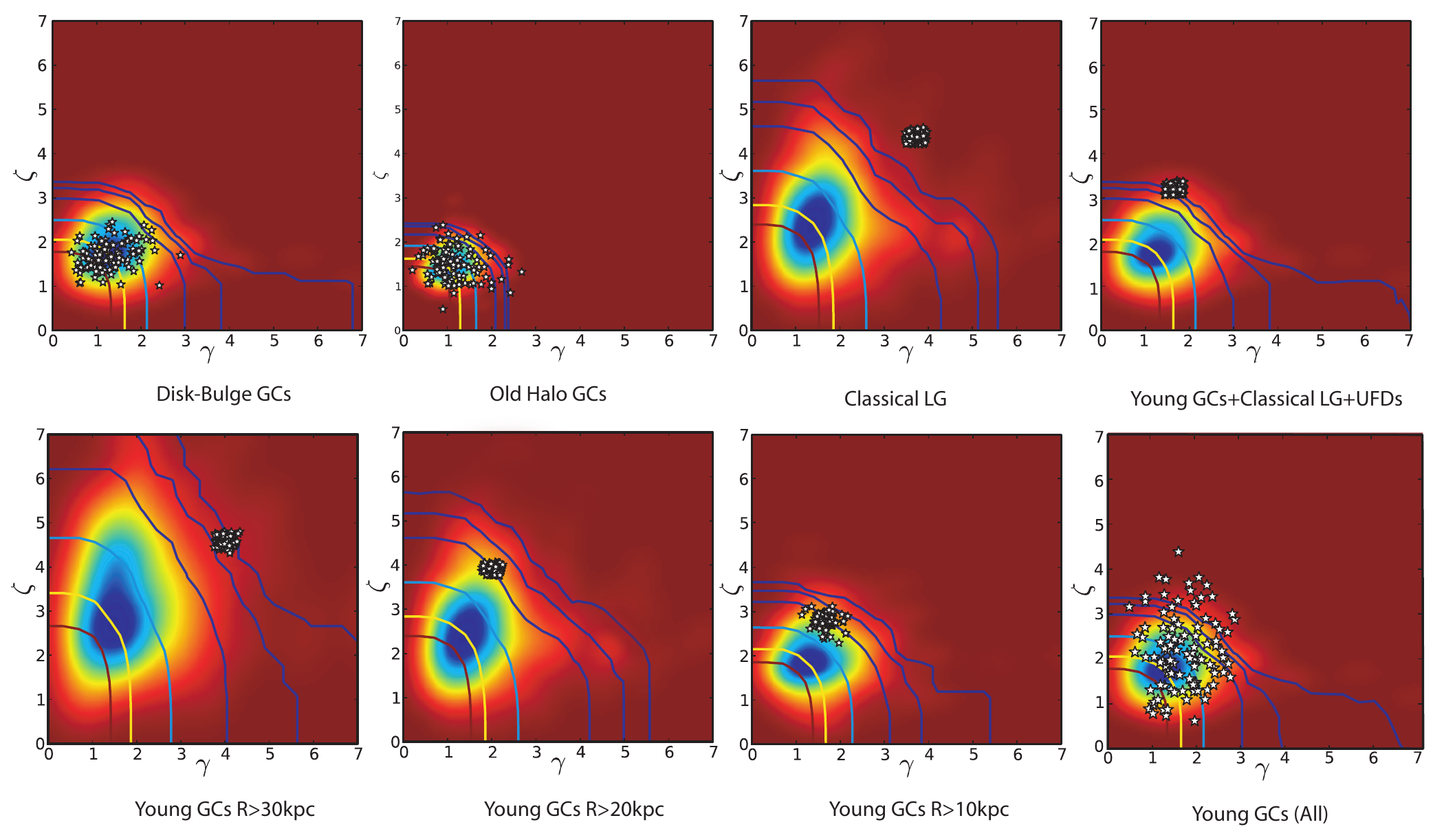}
\caption{A ($\gamma$, $\zeta$) plot (see text for details) describing the spherical distribution of realisations of the observed GC populations as indicated (star symbols, a random sample of 100 realizations of the 10000 computed). The underlying filled color contours show the probability density function arising from repeat realisations of an isotropically distributed population. The solid lines show the locus over which (75\% (red), 50\%(yellow), 20\%(light blue), 5\%(dark blue), 2\% and 1\%) of the sample is of greater $\gamma$ and $\zeta$.}\label{figure:gammazeta}
\end{center}
\end{figure*}

The results of this comparison are seen in Figure \ref{figure:gammazeta}. Here the shaded regions show the probability distribution function of the isotropic sample (realised 10$^4$ times). The star symbol in each panel shows the peak of the probability density function for the observed GC sample (simulated 10$^4$ times to account for the distance uncertainty to each object). In the case of the OH and BD GCs, the spatial distribution of the observed sample is indistinguishable from that of an isotropic distribution of the same size. With no restriction on the inner radius of the YH GC sample, there is a suggestion from the offset of the probability density function peaks that the YH sample is less well described by an isotropic parent distribution (see Figure \ref{figure:gammazeta} lower right panel).

To quantify the hypothesis that the spatial distribution of the observed sample is derived from an isotropic parent distribution we determine the contour lines seen in Figure \ref{figure:gammazeta}. These contour lines enclose regions of the diagram which contain a given percentage of the isotropic sample. The percentage of times the centroid of the observed probability density function is replicated by the isotropic simulated sample for each GC group is given in Table \ref{table:lsq}.

The validity of a spatially isotropic parent population does not change significantly for the OH and DB GC groups as we constrain the galactocentric radius of the sample. In the case of the OH GC group, 19 clusters are of $R_{GC} > 10$ kpc and 6 are $R_{GC} > 20$ kpc, however, these samples are consistent with an isotropic parent distribution in 54\% and 56\% of trials respectively. In the case of the YH GC group, however, the hypothesis of a spatially isotropic parent distribution becomes increasingly unlikely as we restrict the sample to more distant objects. As seen in the $\gamma$-$\zeta$ plane (Figure \ref{figure:gammazeta}) the YH GCs are increasingly located at larger $\gamma$ and $\zeta$ with increasing galactocentric distance. This is indicative that the spherical distribution of the normal vectors to the best-fitting plane for the population is becoming increasingly confined to a preferred direction. That is to say, the YH GCs are seen to be increasingly confined to a planar alignment as we progress to larger galactocentric distances. In Table \ref{table:lsq} we summarize our findings regarding the orientation and rms thickness of the plane that best matches to the observed sample.

We now turn our attention to the spatial distribution of the satellite galaxies of the MW. Figure \ref{figure:gammazeta} shows the corresponding $\gamma$-$\zeta$ plot for the 11 most luminous satellites of the Milky Way (i.e.\ identical to \citet{Metz07}). The significance, orientation and thickness of a plane to the classical dwarf galaxy sample is given in Table \ref{table:lsq}. Our results may be compared to those of \citet{Metz07} who find $\ell =158\mathring{.}2$, $\textit{b} =-11\mathring{.}9$, and a rms thickness, $\Delta$=32.6 kpc. Figure \ref{figure:gammazeta} (center) shows our findings with the inclusion of 10 recently discovered low-luminosity MW satellites (replicating the sample of \citet{Metz09}). Here our results may be compared to those of \citet{Metz09} who find $\ell = 159\mathring{.}7 \pm 2\mathring{.}3$, $\textit{b} = -6\mathring{.}8 \pm 2\mathring{.}3$, and $\Delta$=24.9$\pm 1.1$ kpc. It may be argued that on the basis of the results of \citet{Besla10} that the Magellanic Clouds (MCs) be considered separately since their proper motions suggest they are on their first passage by the MW. On the other hand, since the MCs lie on a plane common to the remaining 9 'classical' satellites we propose that if the MCs are on their first infall that they are further evidence in favor of preferred accretion along a common plane. In both of the above samples we recover the significance, orientation and thickness of the planar distributions discussed in the literature. 

It is apparent from Table \ref{table:lsq} that the orientation and thickness of the plane that describes the spatial distribution of the YH GCs at $R_{GC} > 10$ kpc is indistinguishable from that of the classical MW dwarf galaxies and the recently discovered `ultra-faint' dwarf galaxies. For this reason we combine these three samples to define a common plane of satellites (PoS).

\begin{table}
\caption{The results of our determination of best fitting plane to the distribution of objects as described in $\S$\ref{section:spatial_dist}.}
\begin{center}
\begin{tabular}{lcccccc}
\tableline\tableline
 GC & $R_{GC}$ & Number &$\ell$& $\textit{b}$& rms thickness\tablenotemark{a}  & Significance\tablenotemark{b}\\
 Group & (kpc) & clusters &($^{\circ}$)& ($^{\circ}$) & (kpc)  & (\%)\\
\tableline
OH & All & 56 & \ldots & \ldots &\ldots & 54\\
OH & $>20$ & 6 & \ldots & \ldots &\ldots & 56\\
BD & All & 28 & \ldots & \ldots &\ldots & 53\\
YH & All & 30 & 144 $\pm$ 6 & $-13 \pm 5$ & $28 \pm 5$ & 64\\
YH & $>10$ &23 & 156 $\pm$ 6 &$-8  \pm 5$  & $24 \pm 4$ & 95\\
YH & $>20$ &10 & 169 $\pm$ 5 & $-9   \pm 6$  & $23.2 \pm 2.7$ & 95\\ 
YH & $>30$ &7 & 162 $\pm$ 5 & $-8   \pm 7$  & $18.5 \pm 2.3$ & 99.2\\ 
CS\tablenotemark{c}  & All & 11 & $156.8 \pm 2.7$ & $-11.6   \pm 2.9$ & $19.3 \pm 2.1$ & 99.1\\ 
CS+UFD\tablenotemark{d} & All & 21 & $168 \pm 4$ & $-13   \pm 3$&$25 \pm 3$ & 98\\ 
YL02\tablenotemark{e} & All & 7 & $183 \pm 15$ & $-3 \pm 9$&$19 \pm 4$& 96\\ 
\tableline
\end{tabular}
\tablenotetext{a}{Calculated from repeat random simulations of the distance uncertainties for each object. See text for details.}
\tablenotetext{b}{Percentage of times the observed result is not replicated drawing from an isotropic population.}
\tablenotetext{c}{The 11 most luminous (`classical') satellite galaxies of the Milky Way.}
\tablenotetext{d}{As above with the inclusion of 10 low luminosity dwarf galaxies. See text for details.}
\tablenotetext{e}{The cluster sample of \citet{Yoon02}.}
\end{center}
\label{table:lsq}
\end{table}

\begin{figure*}
\begin{center}
\includegraphics[scale=1, angle=0]{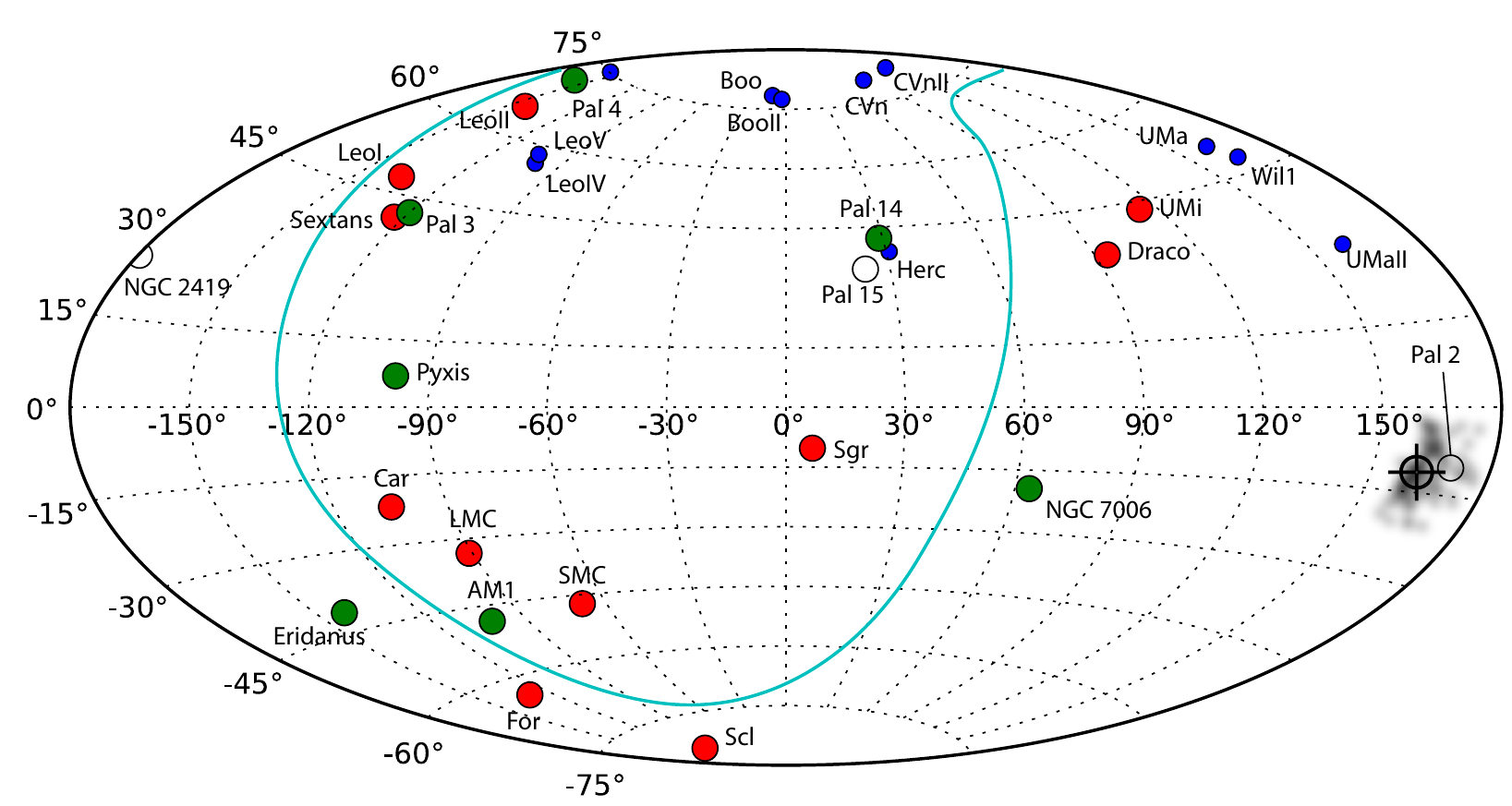}
\caption{The plane of satellites (solid line; derived from the young halo GC population of galactocentric distance greater than 10 kpc; see Table \ref{table:lsq}) seen on an Aitoff projection in Galactic coordinates. Red, green filled circles represent the 11 'classical' luminous satellite galaxies of the Milky Way and the YH globular clusters ($R_{GC}>30$ kpc) respectively. The open circles show the positions of outer OH clusters. The smaller blue points show the positions of the 'ultra-faint' dwarf satellites. Note that this figure is a two dimensional projection of the three dimensional distribution of satellites. For this reason the distance from the line (which is the plane of satellites seen at infinity) does not correspond to distance from the plane. The region of greyscale on the right of the figure shows the probability density function of the normal to this plane from repeated random realisations of the set of objects to account for uncertainties in their distance. The cross hair denotes the position of the highest likelihood position of this normal to the plane.}\label{figure:PoS}
\end{center}
\end{figure*}

Figure \ref{figure:PoS} shows the spatial distribution on the sky of the classes of object we have considered. The PoS, as defined above, is shown as the solid line. The grayscale probability density function seen at $\ell \sim 156^{\circ}$ shows the positions of the normal to this plane in $10^4$ random realisations of the input catalogue distance uncertainties.

\begin{figure*}
\begin{center}
\includegraphics[scale=0.85, angle=0]{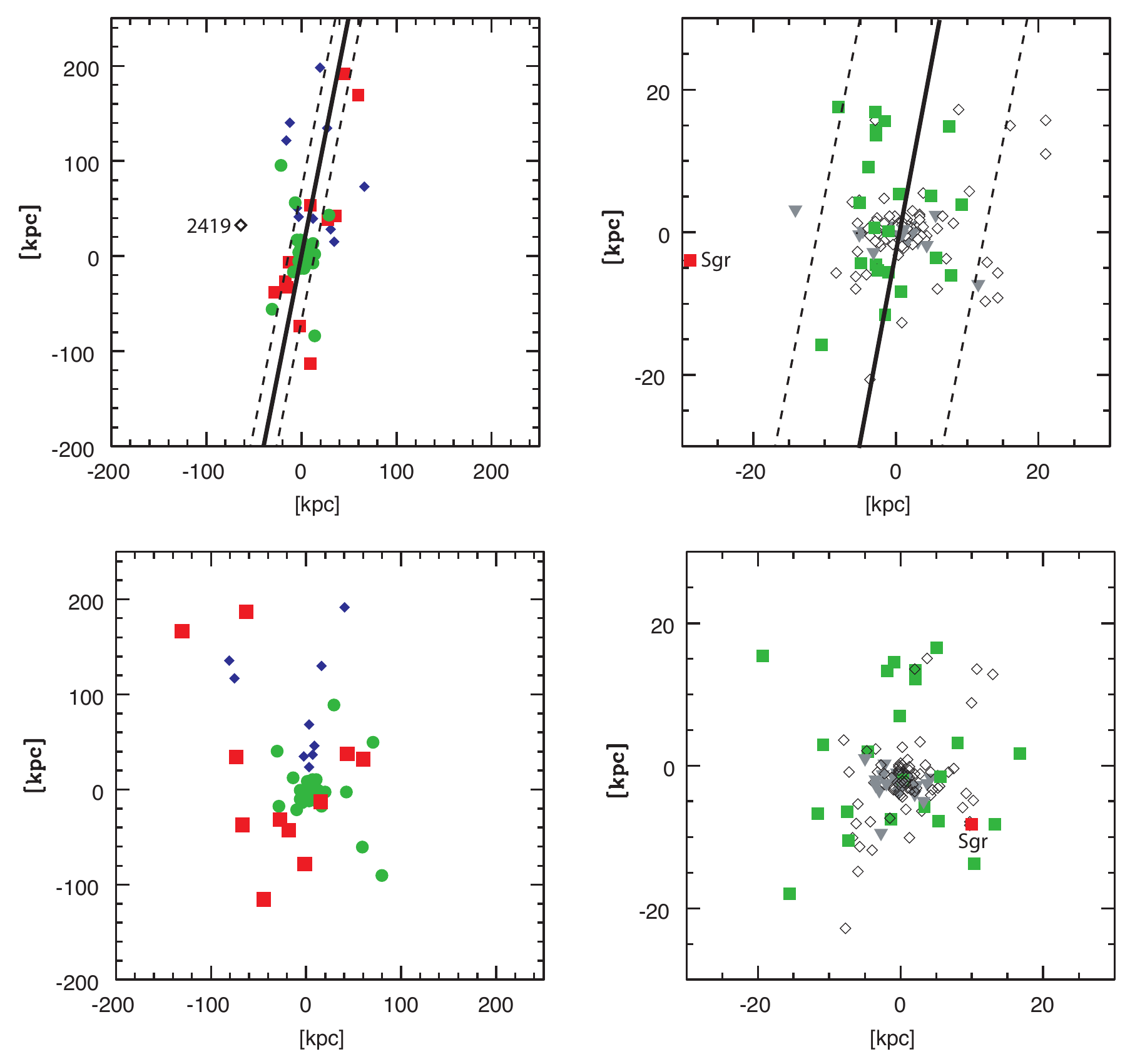}
\caption{{\bf{Top left}}: The PoS described by the young halo GC population of galactocentric distance greater than 10 kpc (see Table \ref{table:lsq}) as seen edge-on to the plane (solid line, surrounded by dashed lines showing the rms thickness, viewed from $\ell$, $\textit{b}$ = ($160^{\circ}$, $0^{\circ}$)). The symbols are colored the same as in Figure \ref{figure:PoS}. {\bf{Top right}}: a zoom in on the inner halo, showing in green the YH clusters, as solid triangles the disk-bulge GCs, and as open diamonds the old halo GCs. {\bf{Bottom left \& right}}: the PoS seen face-on (viewed from $\ell$, $\textit{b}$ = ($50^{\circ}$, $0^{\circ}$)).}\label{figure:PoS_slices}
\end{center}
\end{figure*}

The top left panel of Figure \ref{figure:PoS_slices} shows a section through the vicinity of the MW as seen at a viewing angle edge on to our derived PoS. We note that the OH GC NGC 2419 (the open diamond in the figure) resides well away from this plane at a distance of 83 kpc, 3.4 times the thickness of the PoS. There is increasing evidence that NGC 2419 is an accreted object. Specifically, \citet{Cohen10}, argue that due to its internal [Fe/H] spread this cluster is the former nucleus (or nuclear star cluster) of a disrupted dwarf. Potentially therefore NGC 2419 had been accreted away from the plane apparently preferred by the vast majority of the MW's accreted objects.

Our findings reinforce those of \citet{Yoon02} from a sample of GCs restricted to $R_{GC} < 20$ kpc. In that study, these authors isolated a set of young halo clusters based on metallicity and RR Lyrae pulsation period (as discussed in \S\ref{Section:partitionGCs}). Their set of seven young GCs (their `group II-b' ) describe a plane orientated towards $\ell =183^{\circ} \pm 15^{\circ}$, $\textit{b} =-3^{\circ} \pm 9^{\circ}$ (at 96\% significance compared to an isotropic distribution), which is consistent with the PoS derived here. The PoS is therefore,  persistent in the spatial distribution of the MW's YH GCs to the inner galaxy.

Although we here use YH GCs as a tracer of accreted systems, it is expected that accreted systems will deposit both YH and OH clusters. As pointed out by \citet{DaCostaArmandroff95} the Sgr dwarf is contributing to both groups. Furthermore, \citet{Mackey04} propose that 15-17\% of the OH GCs were contributed via accretions. The fact that our analysis had not shown any radial anisotropy in the distribution of OH GCs is perhaps a consequence of the smaller fraction of accreted systems in this grouping compared to the YH GCs.

We have shown that the YH GC population at galactocentric distances exceeding 10 kpc exhibits the {\it{same}} orientation (within uncertainties) as that defined by the MW satellites. This argues for a strong causal link between the origins of the outer halo GCs and the MW satellites. In the next section we discuss the origins of the planar arrangements of dwarf galaxies around the MW and M31, and by direct association, the origin of the outer halo globular clusters.


\section{The satellites of the Milky Way and M31: cosmological context}

Three scenarios are presented in the literature regarding the development of the spatial distribution of satellites: that the satellites are in-situ tracers of a prolate dark matter halo; that they are remnants from the break up of a large progenitor at early times (i.e.\ the satellites are tidally disrupted galaxies -- TDGs);  or they are tracers of collapse along filaments of large scale structure (LSS). 
Cosmological simulations of dark matter halos shows that under $\Lambda$CDM cosmology halos are predominantly prolate \citep{Libeskind05, Zentner05}, however, the degree to which the dark matter halo would have to be prolate to accommodate the observed anisotropy around the MW and M31 would be extremely atypical \citep{Kroupa10}. Furthermore, the debris stream of the disrupting Sagittarius dwarf can not be modelled under the influence of such an extremely prolate dark matter halo \citep{Law09, Prior09b, Keller08, Keller09}.

\citet{Li08} present a scenario in which all satellites fall in to the MW halo in 1--2 groups and demonstrate that such an accretion history could account for the current satellite anisotropy. \citet{DOnghia08} develop a similar scenario, reminiscent of the groups of MW satellite galaxies of \citet{Lynden-Bell76, Lynden-Bell82, Kunkel76}, in which the groups formed in LMC-like dark matter halos.  However, such a scenario has a number of problematic features as discussed in \citet{Kroupa10}, the frequency of such groups in the local volume is low; the group would have been comprised predominantly of dwarf spheroidal galaxies and this does not match to the morphology-density relation; and the group identity of satellites after infall is expected to be short lived \citep{Klimentowski09}.

\citet{Metz07} present a scenario in which the satellites arise from the tidal fragmentation of a large galaxy at an early epoch. Under this scenario, the kinematics of the TDGs is retained from formation as a rotationally supported disk. \citet{Metz08,Metz09} confirm this: the inferred angular momentum vectors for the majority of classical MW satellites are directed within a 30$^{\circ}$ region on the sky. 

\citet{Libeskind05, Libeskind10} present cosmological simulations that show that substructures stream through LSS filaments that feed MW-sized halos. This accretion occurs from preferred directions on the sky and is coherent on scales from 1 Mpc to 20 kpc. From simulated systems that contain at least 11 luminous satellites (i.e. comparable to the MW system) \citet{Libeskind10} find 30\% possess satellites with orbital angular momenta aligned to a similar degree as the MW system. \citet{Lovell10} find a similar result, namely that quasi-planar distributions of coherently rotating satellites arise naturally in the Aquarius simulations as subhalos fall in along the central spines of LSS filaments\footnote{However we note that \citet{Kroupa10} argues this is a selection effect citing that the MW-like galaxies with at least 11 luminous satellites account for 1.4\% of simulated galaxies in \citet{Libeskind10} and hence that only 0.4\% (i.e.\ 30\% of 1.4\%) of all existing MW-like $\Lambda$CDM halos would host a MW-type galaxy with an appropriate satellite spatial distribution.}.

\begin{figure*}
\begin{center}
\includegraphics[scale=0.9, angle=0]{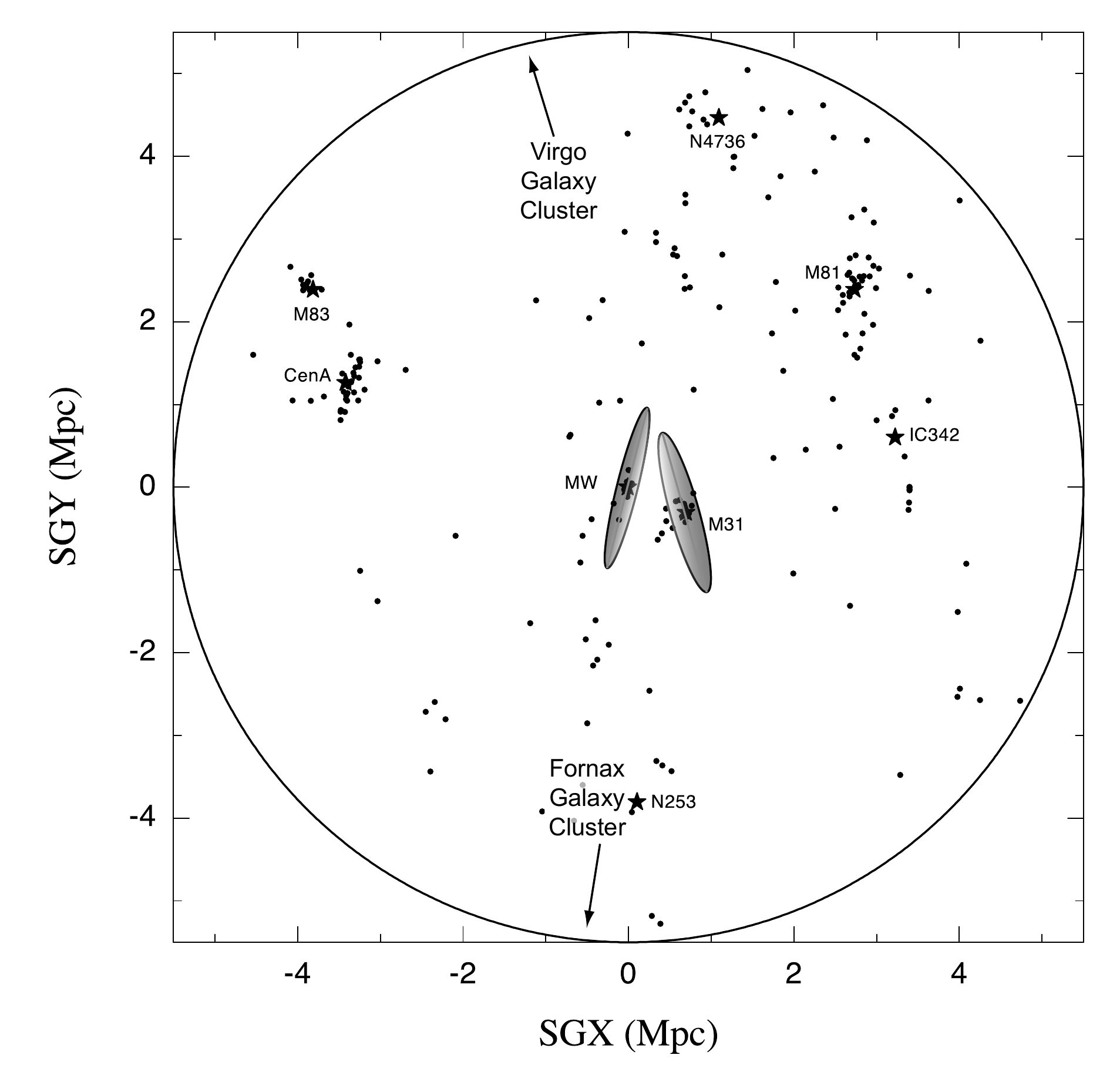}
\caption{The PoS of the Milky Way and M31 seen in the context of the supergalactic coordinate frame of reference. The shading of the ellipses represents orientation, darker is closer. In this X-Y projection the PoS of both galaxies are nearly side-on and generally directed along an axis joining nearby major mass concentrations in the form of the Virgo and Fornax galaxy clusters (15 and 12 Mpc  distant from the Milky Way respectively).}\label{figure:SGxy}
\end{center}
\end{figure*}

\begin{figure*}
\begin{center}
\includegraphics[scale=0.9, angle=0]{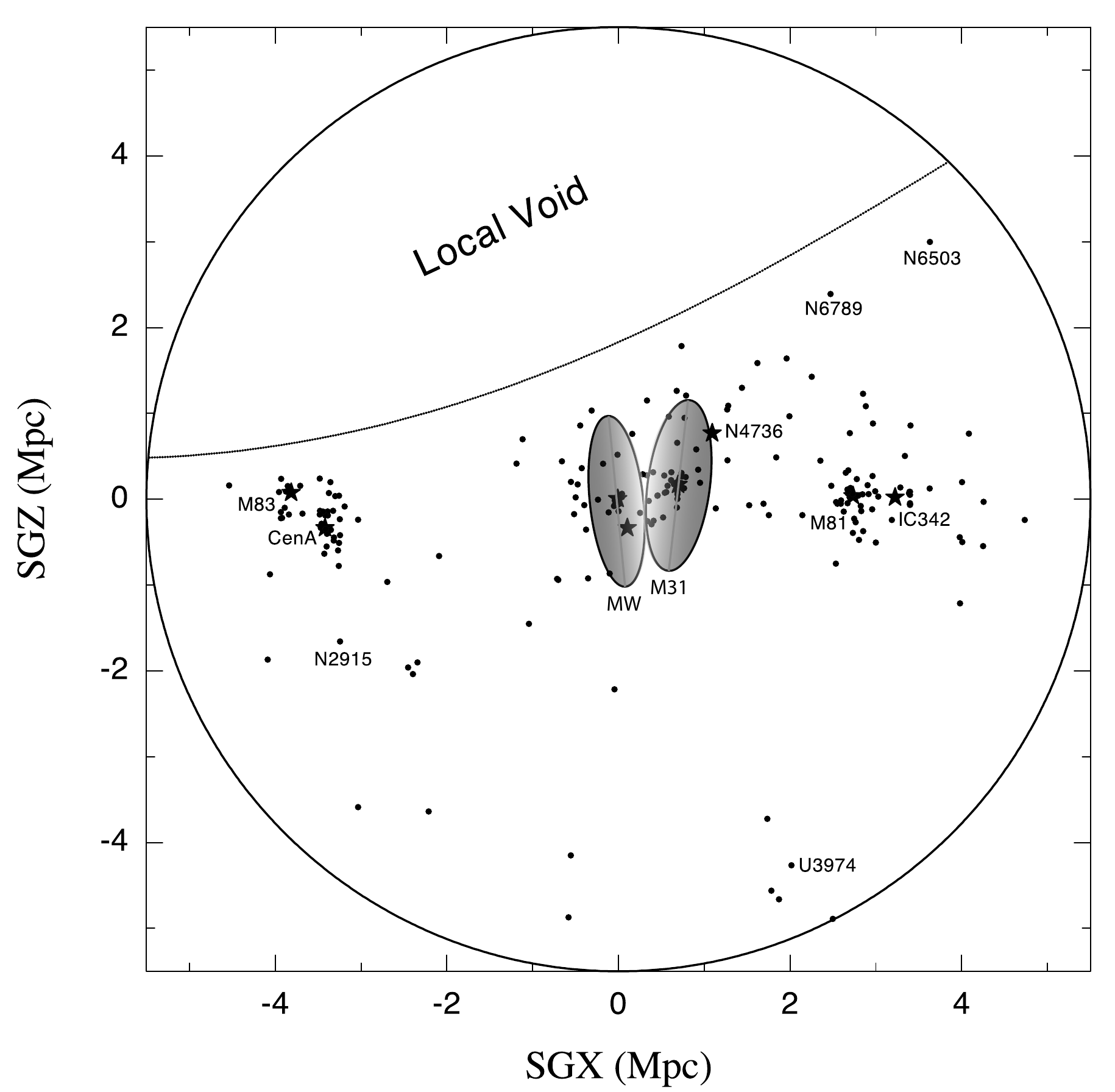}
\caption{As in Fig. \ref{figure:SGxy} but in the X-Z section through the supergalactic plane.}\label{figure:SGxz}
\end{center}
\end{figure*}

In Figures \ref{figure:SGxy} and \ref{figure:SGxz} we show the orientation of the MW and M31 PoS in supergalactic coordinates (see \citeauthor{deVaucouleurs91} \citeyear{deVaucouleurs91} for definition). The supergalactic plane is defined by the distribution of galaxies $\sim$ few Mpc about the MW. Within this volume galaxies are confined to a $\sim 2$ Mpc thick plane (seen side on in Figure \ref{figure:SGxz}).  Seen from above (Fig. \ref{figure:SGxy}) the MW and M31 PoS are both highly inclined with respect to the supergalactic plane. As discussed in \citet{Metz07} there is no apparent spatial connection between the MW and M31 PoS as the M31 PoS is inclined at $\sim 55^{\circ}$ to the MW PoS. It is, however, notable that the PoS of both galaxies is well aligned with an axis between the two major mass distributions within 20 Mpc: the Virgo and Fornax clusters \citep{Karachentsev03}. This was remarked upon by \citet{Navarro04} in their study of the galaxies of the local supergalactic plane\footnote{However, we note that \citet{Yang06, Agustsson06} find that this alignment does not appear to extend to scales less than the typical virial radius.}. Here they found that not only are the satellites of nearby early-type galaxies preferentially confined to polar planes relative to their disks but that the galaxy disk is orientated along the major axis of surrounding LSS material. This is also the case for the MW and M31, where as seen in Fig.\ \ref{figure:SGxy} and \ref{figure:SGxz}, the PoS are highly inclined to the stellar disks of their hosts. 


\section{Discussion: the origin of outer halo globular clusters}
It is our thesis that the observed affinity between the planar distribution of outer halo GCs and that of the MW satellites implies a common origin. In this scenario, the outer halo GCs are accreted to the MW as part of galaxies from the surrounding LSS. They enter as entities residing in dark matter subhalos in LSS filaments that stream into the MW dark matter halo. Upon entry, potential subhalos may deliver the GCs from their natal dark matter subhalos via tidal disruption and dispersal of dwarf galaxies (akin to the simulations of \citeauthor{Libeskind10} \citeyear{Libeskind10}) or from the early tidal disruption of a moderate mass galaxy (like that envisaged by \citeauthor{Kroupa10} \citeyear{Kroupa10}). We discount the possibility of "free-floating" GCs as this would imply dark matter dominance and in the case of at least one of the outer halo GCs (Pal 14; \citeauthor{Jordi09} \citeyear{Jordi09}) there is no evidence for dark matter.

We see in the spatial distribution of the outer halo GCs evidence of their accretion origin. This is a direct observation of the importance of accretion to galaxy formation as inferred from the properties of the MW GC system in the seminal work of \citet{SZ}. The only direct observations of this process of globular cluster accretion are in the disrupting Sagittarius dwarf (delivering at least five GCs into the MW halo: \citeauthor{DaCostaArmandroff95} \citeyear{DaCostaArmandroff95}; \citeauthor{MartinezDelgado02} \citeyear{MartinezDelgado02}; \citeauthor{Law10} \citeyear{Law10}), perhaps in the case of the putative Canis Major dwarf \citep{Martin04} and in the globular clusters of M31 associated with tidal stellar streams in the vicinity of M31 \citep{Mackey10}.

Much corroborating evidence is presented for this conclusion is presented in the literature. A particularly distinctive feature of the YH GCs is their distribution of core radii (\citeauthor{Mackey04} \citeyear{Mackey04}; see also \citeauthor{MackeyvdB05} \citeyear{MackeyvdB05}). The core radii ($r_c$) of the YH GCs shows a very long tail to very large radii (all but one GC with $r_c > 9$ pc resides in the \citeauthor{Mackey04} grouping of GCs associated with the outer halo GCs). Furthermore, the $r_c$ distribution of these GCs shows no statistical difference to that observed from a compilation of GCs from the LMC, Fornax and Sagittarius dwarf galaxies \citep{Mackey03b}. Together with the similarity in morphology of the horizontal branch between the external GCs and the YH GC grouping, this evidence leads \citet{Mackey04} to propose that all the YH GCs are accreted.

The implications of the size distribution for an accreted origin of the outer halo GCs are elucidated in the study of \citet{Hurley10}. In this study, the authors use N-body simulations of star clusters in a tidal field to investigate the conditions required to produce and sustain a GC of large core radius. \citeauthor{Hurley10} show that clusters may be born with a range of $r_c$ governed by how much the cluster fills its initial tidal radius. When a cluster completely fills its natal tidal radius an extended GC can result ($r_c > 10$ pc). Conditions for a cluster to completely fill its tidal radius at birth are optimal in regions where background tidal forces are low: such conditions are best satisfied at large distances from MW-like galaxies and in dwarf systems (see discussion in \citeauthor{Elmegreen08} \citeyear{Elmegreen08} and \citeauthor{DaCosta09} \citeyear{DaCosta09}). This again argues for the accretion of the outer halo GCs to the MW.

The timing of the delivery of the outer Young Halo GCs through accretion is not constrained by the physical parameters of the GCs. \citet{Gnedin97} describe the stability of the MW's GCs against two-body relaxation, tidal truncation, and tidal shocks due to passage through the disk and due to close proximity to the bulge. With the exception of two clusters (Pal 1 and Pal 13), GCs of R $> 10$ kpc are expected to be long-lived, with lifetimes of between 5 and 100 Hubble times. Hence, by this criterion, these systems could have been accreted into the MW at any stage over the last Hubble time. We can ask the question: how many satellites of a given mass are required to contribute the observed (conservative) 22 Young Halo GCs at R $>10$ kpc. Given the specific frequency of GCs ($S_{N}$)as a function of host galaxy luminosity (first introduced by \citeauthor{Harris81} \citeyear{Harris81}, and e.g.\ discussed recently by \citeauthor{Georgiev10} \citeyear{Georgiev10}), we find this would require for example, approximately 2 Magellanic-like (M$_{V} \sim -18$ with GC $S_{N} \sim 1$) systems or 22 systems with Sculptor-like luminosities (M$_{V} \sim -11$ with GC $S_{N} \sim 70$). If we then consider that the number of accreted clusters is likely to be supplemented by 10-12 OH clusters \citep{Mackey04} we estimate that the MW may have experienced mergers with 3 Magellanic-like to 30 Sculptor-like luminosity systems over the MW's lifetime\footnote{When considering the number of mergers it should be noted that there is almost a factor of ten variation exhibited in the $S_{N}$ at given galaxy luminosity}. Such deposition would contribute around half of the mass of the stellar halo. In order to contribute stars that are of sufficiently low metallicity (and sufficiently high [$\alpha$/Fe]) to match those found in the halo, such accretion must have taken place early on in the chemical evolution of the accreted galaxies.

\section{Conclusions}

In this study we have partitioned the Milky Way's globular cluster system into groups according to metallicity, horizontal branch morphology, and the stellar pulsation properties of their RR Lyrae population. The grouping broadly align with relative ages of the constituent clusters. When partitioned in this way the Milky Way's young halo globular clusters (of galactocentric radius greater than 10 kpc) are anisotropically spatially distributed and the distribution can be described by a plane $24 \pm 4$ kpc in thickness (rms). The normal to this plane is directed towards ($\ell$, $\textit{b}$) = ($156^{\circ}\pm 6^{\circ}$, $-8^{\circ} \pm 5^{\circ}$). In contrast the old halo and bulge-disk groups are consistent with an isotropic spatial distribution. The plane defined by the young halo population is statistically indistinguishable from that described by the satellite galaxies of the Milky Way ($\ell$, $\textit{b}$ = $166^{\circ} \pm 4^{\circ}$, $-13^{\circ} \pm 3^{\circ}$). We argue therefore that the young halo globular clusters share the same origin as that of the system of satellite galaxies about the Milky Way.

Simulations of the satellite galaxies of Milky Way-like systems have shown that the planar arrangement of satellites arises naturally from accretion of subhalos along filaments of the surrounding large scale structure which are directed towards the host halo. Once accreted, the globular clusters of the subhalos are delivered to the Milky Way halo by tidal disruption. 

The finding of a common spatial anisotropy amongst the Milky Way satellites and the young halo globular clusters is an observation of the accretion origin of the outer globular clusters. It is a direct confirmation of numerous lines of evidence that have shown the outer halo globular clusters to possess individual properties that are shared most closely with those found in the globular cluster systems of low mass galaxies. 

\acknowledgments
This paper arose from discussions at the {\it{Galactic Studies with the LAMOST Surveys}} workshop held at the Kavli Institute for Astronomy at Peking University, July 2010. SK \nocite{Keller07} and DM would like to express their gratitude to the organisers for providing a fruitful meeting. SK \& GDaC's research has been supported in part by the Australian Research Council through Discovery Project Grant DP0878137. DM acknowledges the financial support from an Australian Research Fellowship (DP1093431) from the Australian Research Council.

\clearpage

\clearpage

\clearpage

\clearpage

\clearpage

\clearpage

\end{document}